\documentclass[twocolumn]{aastex63}
\received{2020 May 15}
\revised{2020 June 16}
\accepted{2020 June 22}
\submitjournal{ApJL}
\shorttitle{Steady-flow driver}
\shortauthors{Karampelas et al.}
\graphicspath{{./}{figures/}}

\begin{document}

\title{Generating transverse loop oscillations through a steady-flow driver}

\correspondingauthor{Konstantinos Karampelas}
\email{kostas.karampelas@kuleuven.be}

\author[0000-0001-5507-1891]{Konstantinos Karampelas}
\affiliation{Centre for mathematical Plasma Astrophysics, Department of Mathematics, KU Leuven,\\ Celestijnenlaan 200B bus 2400, B-3001 Leuven, Belgium }

\author[0000-0001-9628-4113]{Tom Van Doorsselaere}
\affiliation{Centre for mathematical Plasma Astrophysics, Department of Mathematics, KU Leuven,\\ Celestijnenlaan 200B bus 2400, B-3001 Leuven, Belgium }

\begin{abstract}

In recent years, the decay-less regime of standing transverse oscillations in coronal loops has been the topic of many observational and numerical studies, focusing on their physical characteristics, as well as their importance for coronal seismology and wave heating. However, no definitive answer has yet been given on the driving mechanism behind these oscillations, with most studies focusing on the use of periodic footpoint drivers as a means to excite them. In this paper, our goal is to explore the concept of these standing waves being self-sustained oscillations, driven by a constant background flow. To that end, we use the PLUTO code, to perform $3$D magnetohydrodynamic simulations of a gravitationally stratified straight flux tube in a coronal environment, in the presence of a weak flow around the loop, perpendicular to its axis. Once this flow is firmly set up, a transverse oscillation is initiated, dominated by the fundamental kink mode of a standing wave, while the existence of a second harmonic is revealed, with a frequency ratio to the fundamental mode near the observed ones in decay-less oscillations. The presence of vortex shedding is also established in our simulations, which is connected to the ``slippery" interaction between the oscillator and its surrounding plasma. We thus present a proof-of-concept of a self-oscillation in a coronal loop, and we propose it as a mechanism that could interpret the observed decay-less transverse oscillations of coronal loops.

\end{abstract}


\section{Introduction} \label{sec:intro}

The theory of magnetohydrodynamic waves in a simple cylindrical flux tube \citep{zajtsev1975, edwin1983wave} has been used to describe the different modes expected in structures commonly found in the solar atmosphere. Observations by the Coronal Multi-channel Polarimeter, the Solar Dynamics Observatory, and Hinode spacecraft have already proved the ubiquity of such transverse perturbations along coronal loops, prominence threads, and greater areas of the corona \citep[e.g.][]{okamoto2007, tomczyk2007, mcintosh2011}, as well as raising arguments over the magnitude of the estimated energy carried by such waves  \citep[e.g.][]{depontieu2007, morton2016ApJ}. The latter is of great importance since wave energy dissipation is one of the possible coronal heating models \citep{Arregui2015review}.

In particular, kink oscillations of solar coronal loops have been intensively
studied ever since their observation \citep{aschwanden1999, nakariakov1999}. The most common explanation for the nature of these waves is that they are standing kink modes of coronal loops \citep{tvd2008detection}. The observed kink oscillations are categorized into two different groups: the large-amplitude decaying oscillations \citep[e.g.][]{aschwanden1999}, and the small-amplitude decay-less oscillations \citep[e.g.][]{nistico2013}. 

Decaying oscillations usually have amplitudes of a few megameters and are associated with external energetic phenomena \citep{ZimovetsNakariakov2015A&A}. The damping of these oscillations has been attributed to the phenomena of resonant absorption and phase mixing \citep{Ionson1978ApJ, heyvaerts1983,goossens2011resonant} and have been studied both analytically and numerically in 3D MHD setups, where the effects of gravity, radiation, and the Kelvin-Helmholtz instability (KHi) has also been considered \citep[e.g.][]{terradas2008, antolin2014fine, magyar2015, Hillier2019MNRAS}. 

The low-amplitude, decay-less oscillations were first detected in imaging data and spectroscopic data \citep{tian2012, wang2012}, and were later proven to be ubiquitous in active region coronal loops \citep{anfinogentov2013, anfinogentov2015}, making them possible tools for coronal seismology \citep{anfinogentov2019ApJ}. These decay-less oscillations were observed to have a near constant amplitude over the course of many periods, and showed frequencies equal to that of the fundamental standing kink mode \citep{nistico2013}. More recent studies \citep{duckenfield2018ApJ} have also detected the existence of the second harmonic in these waves. In \citet{antolin2016}, these decay-less oscillations have been connected to the line-of-sight effects created by the KHi vortices from impulsively oscillating coronal loops. Another explanation of these oscillations is that they are standing waves initiated by footpoint drivers. A number of recent 3D numerical studies have reproduced the decay-less oscillations, recovering many of their observational characteristics and treating them as candidates for wave heating mechanisms in the solar corona \citep[e.g.][]{tvd2018, mingzhe2019, karampelas2019, karampelas2019amp}. Recent studies have shown that the fundamental kink mode can manifest in oscillating loops even when broadband drivers are considered \citep{afanasev2019, afanasyev2020decayless}, reinforcing the connection between these decay-less oscillations and footpoint driven waves.

In the current study we are going to focus on a different interpretation of these undamped waves that was considered in \citet{nakariakov2016}. In that study, \citeauthor{nakariakov2016} described these oscillations as a self-oscillatory process \citep{jenkins2013PhR}, generated by the interaction of the loops with weak, quasi-steady flows. The main difference of self-sustained oscillations is that the frequency of the oscillations is set by the system itself, rather than the external driver. Here we will show that these oscillations can be excited as a self-sustained process from weak flows around flux tubes in a coronal environment, giving a proof-of-concept for a self-oscillation in a 3D flux tube in a coronal environment.

\begin{figure*}[t]
    \centering
    \resizebox{\hsize}{!}{\includegraphics[scale=0.4]{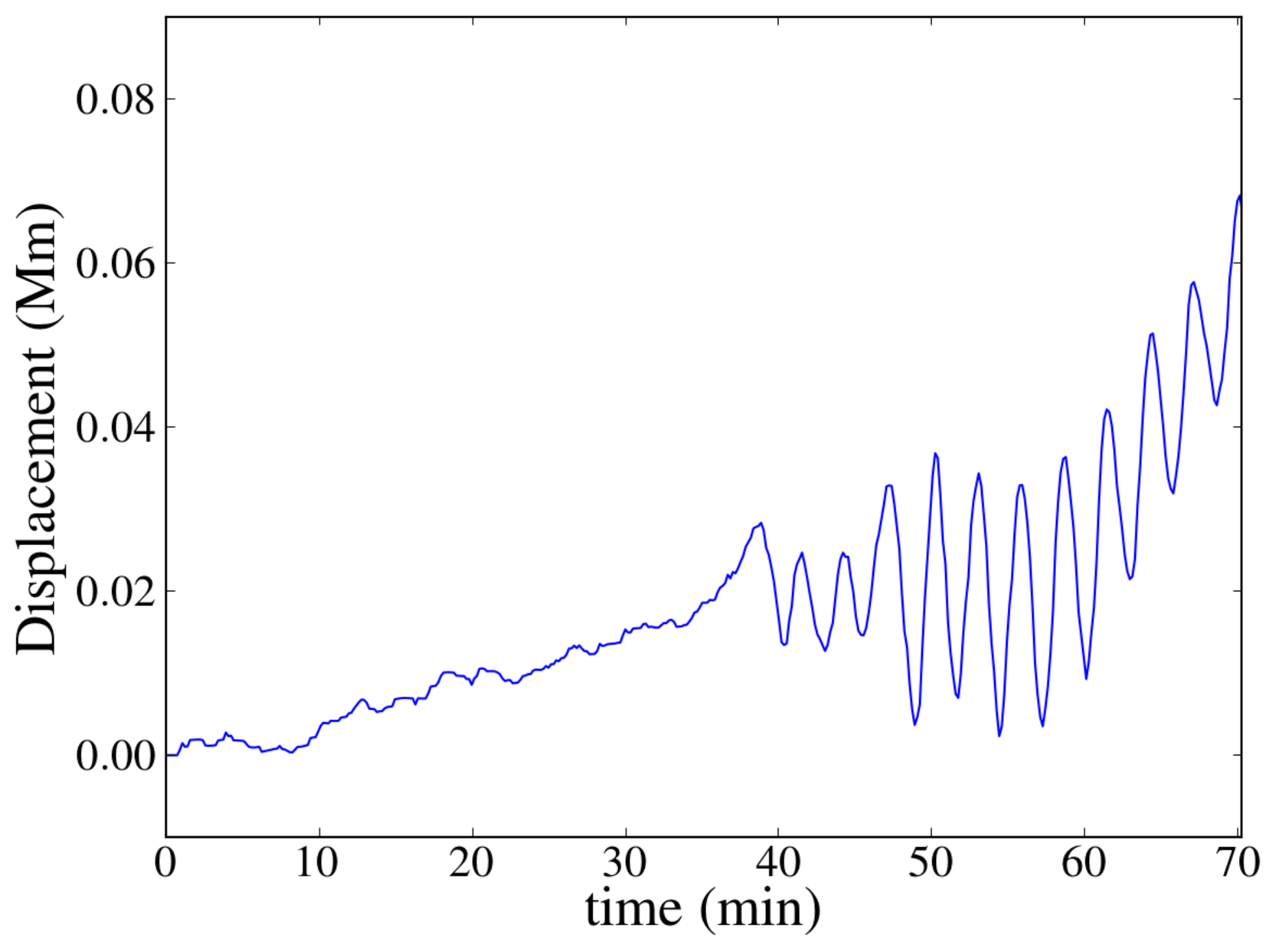}
    \includegraphics[scale=0.4]{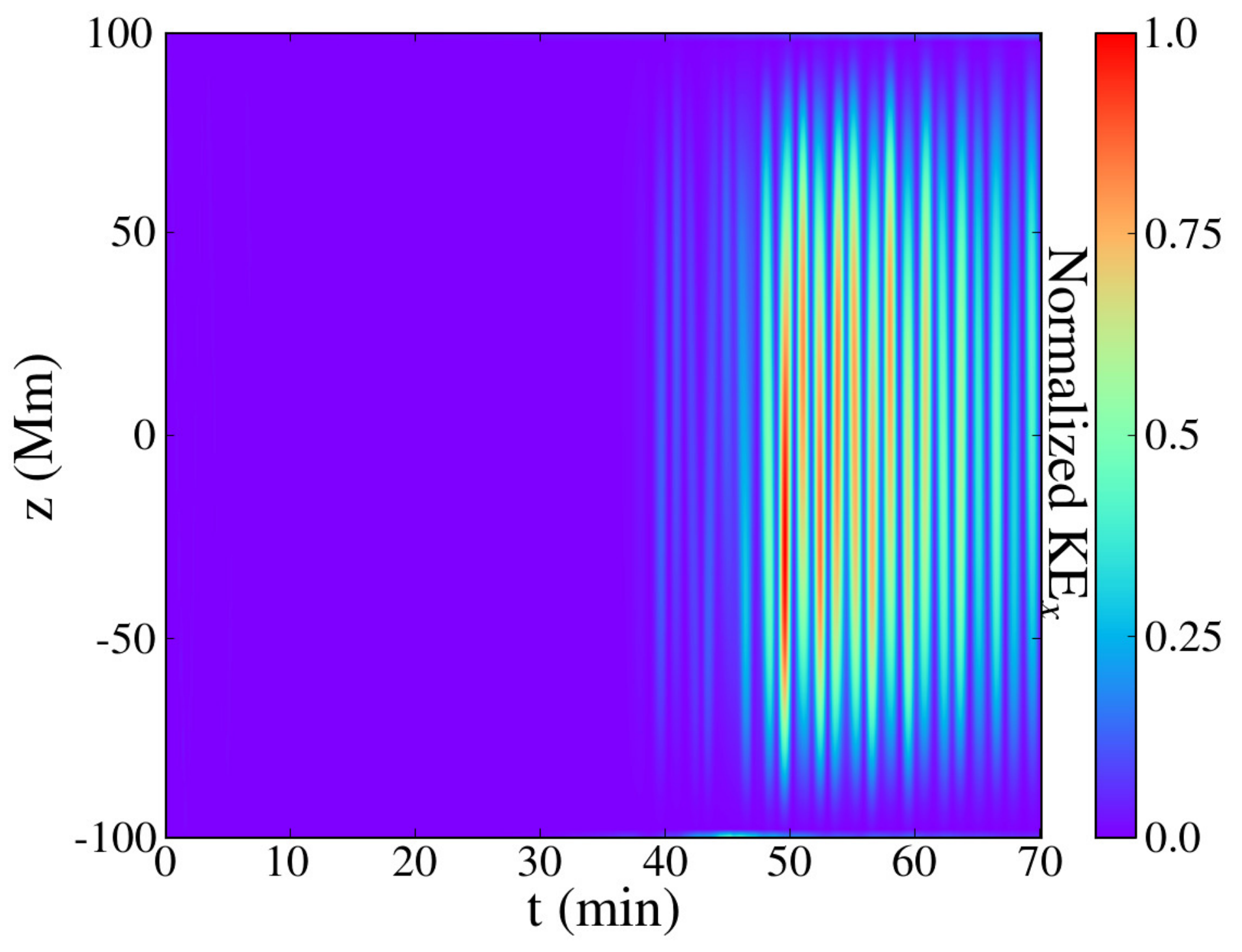}}
    \caption{Left panel: displacement of the apex along the $x$ direction of a coronal loop, excited from a steady background footpoint flow. Right panel: normalized average kinetic energy density inside the loop per height and over time, with only the $\upsilon_x$ component (KE$_x$) considered.}
    \label{fig:osc}
\end{figure*}

\section{Numerical setup} \label{sec:setup}

The main setup consists of a straight flux tube of radius $R=1$\,Mm and length $L=200$\,Mm, consisting of gravitationally stratified plasma, similar to the setup used in \citet{karampelas2019}. The coronal background density at the footpoint is $\rho_e=0.836\times 10^{-12}$\,kg\,m$^{-3}$, three times lower than the loop density at the footpoint ($\rho_i$). The temperature varies across the tube axis (on the $xy$-plane), ranging from $0.9$\,MK inside the loop to $1.35$\,MK outside, while it is constant with height, along the flux tube.  The radial density profile for our models at the footpoints ($z=\pm 100$\,Mm) is given by the relation
\begin{equation}
\rho(x,y) = \rho_e  + (\rho_i - \rho_e)\zeta(x,y), 
\end{equation}
\begin{equation}
\zeta(x,y) = 0.5(1-\tanh((\sqrt{x^2+y^2}/R-1)\,b)),
\end{equation}
where $b$ sets the width of the boundary layer. We consider $b=20$, which gives us an inhomogeneous layer of width $\ell \approx 0.3 R$. The index $i (e)$ corresponds to the internal (external) values with respect to our flux tube.

We consider sinusoidally varying gravity along the flux tube, which models the effects of curvature along an equivalent semicircular loop with major radius equal to $L\pi^{-1}$. We thus have stratification of pressure and temperature along the loop according to the hydrostatic equilibrium,
\begin{equation}
\frac{\partial p_{i,e}}{\partial z}=g\, \rho_{i,e}\, \sin(\frac{\pi z}{L}).,
\end{equation}
where $g=274$\,m\,s$^{-2}$ is the surface gravity of the Sun. We initially consider a magnetic field $B_z$, with values $B_{zi} = 22.7$\,G and $B_{ze} = 22.8$\,G at the footpoints. After letting our system freely evolve for one oscillation period ($\sim 161$ s, see Section 3) at the beginning of the simulations, the magnetic field resettles in an almost straight field parallel to the loop axis, with a slight increase of magnitude toward the apex ($< 0.3$\,G) and with small $B_x$ and $B_y$ magnetic field components ($B_x,B_y\ll 1$\,G). This relaxation also minimizes any unwanted perturbations at the start of the simulation from the redistribution of the magnetic field.

Our setups have domain dimensions of $(x,y,z) = (8,8,200)$ Mm, with a resolution of  $(\delta x,\delta y,\delta z) = (40,40,2000)$\,km  in the $x$, $y$, and $z$ directions respectively. The loop footpoints are placed at positions $z=-100$ and $z=100$\,Mm, while $z=0$ is the location of the loop apex. At the lateral boundaries, we apply outflow (Neumann-type, zero-gradient condition) conditions, which allow waves to leave the domain. To minimize their effect on the dynamics of our loops, the side boundaries are placed at a safe distance from the loop.

At the ``bottom" and ``top" boundaries ($z=-100$ and $z=100$\,Mm), we apply zero-gradient conditions for the pressure, density, and the three components of the magnetic field. The $v_z$ velocity component (along the axis of the loop) is set as antisymmetric, to prevent any outflows from the top and bottom boundaries, where the bases of the loop are located. Inside the loop ($\sqrt{x^2+y^2}\leq R$), the $v_x$ and $v_y$ are set as antisymmetric, to fix the loop endpoints. Outside the loop at $z=-100$\,Mm we set the $v_x=300$\,m\,$s^{-1}$, in order to model a steady flow around the footpoint, originating from the supergranulation motions at the photosphere \citep{RieutordRincon2010}. The $v_y$ component is left free (outflow conditions) in order to let the flow evolve freely along the $y$ direction. At the other boundary ($z=100$\,Mm) outside the loop, both $v_x$ and $v_y$ are left to evolve freely, as this was a way to allow the development of vortices.

All calculations were performed in ideal MHD in the presence of numerical dissipation, using the PLUTO code \citep{mignonePLUTO2012}. We use the second order characteristic tracing method for calculating the timestep, and the finite volume piecewise parabolic method (PPM) with a second order spatial global accuracy.  The solenoidal constraint on the magnetic field is kept with the extended GLM method.

\section{Results} \label{sec:results}

As is described in \citet{jenkins2013PhR}, self-oscillations are processes that can turn a nonperiodic driving mechanism into a periodic signal. One simple example is the excitation of vibrations on a violin string, from the constant movement of a bow slowly moving across the string \citep{goedbloed1995AIPC}. In \citet{nakariakov2016}, it was argued that a loop interacting with a surrounding medium in a ``slippery" fashion, can lead to the development of negative damping, resulting in an oscillation. Using a mechanical analog of a spring pendulum with a weight on a conveyor belt, they applied  a ``slippery" interaction in the form of a friction parameter dependent on the belt's velocity. The resulting motions would quickly turn into an oscillatory pattern, with a frequency identical to the eigenfrequency of the system. If we consider again the bow on a string analog for a coronal loop, slow footpoint flows, like those caused by supergranulation \citep{RieutordRincon2010}, can reach the threshold value for the onset of that negative friction, leading to the onset of a decay-less oscillations.

\begin{figure}[t]
    \centering
    \resizebox{\hsize}{!}{\includegraphics[scale=0.4]{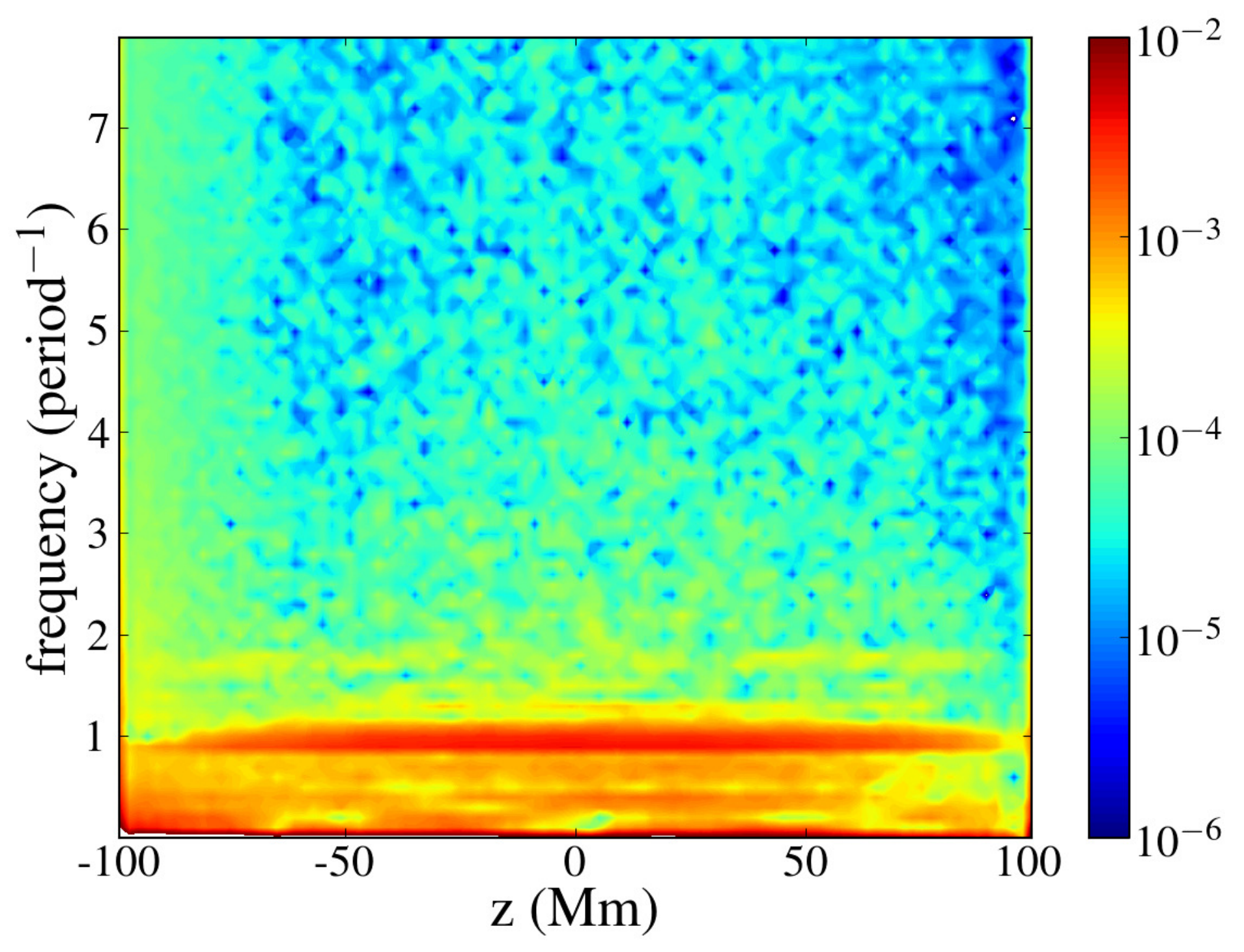}}
    \caption{Power spectral density of the transverse displacement as a function of the $z$ position, caused by a steady background flow at its footpoint ($z=-100$\,Mm). Note the peak at the frequency $1\, P^{-1}$, with period $P \sim 161$ s, due to the excitation of the fundamental kink mode.}
    \label{fig:freq}
\end{figure}

For our simulations, we applied a slow horizontal flow around one of the footpoints, with a velocity of $v_x=300$\,m\,$s^{-1}$, which is the same order of magnitude as those from supergranulations. Due to the straight magnetic field in our setup, the initial perturbation from the initiation of the driver travels along the loop, setting up a horizontal flow at different heights across the loop. Looking at the displacement of the loop center of mass at the apex, on the left panel of Fig. \ref{fig:osc}, we observe that the loop is initially displaced from its initial position, due to that initial flow. Around $t\sim 38$ minutes, an oscillation is initiated with a period $P = 2L/c_k\sim  161$ s ($c_k$ is the kink speed), which is the same as the period of the fundamental standing kink mode, expected for a gravitationally stratified straight flux tube \citep{edwin1983wave, andries2005}. We note that the values of the oscillation amplitude ($0.01$ Mm) are roughly $10$ times smaller than the observed values for decay-less oscillaltions \citep[$\sim 0.1$ Mm, see][]{anfinogentov2015}, and further studies are required in order to explore the viability of this method. However, we prove here for the first time that such a self-sustaining  mechanism is possible in a coronal environment for a $3$D MHD model. 

Looking at the normalized kinetic energy density inside the loop for the $v_x$ velocity (KE$_x$), on the right panel of Fig. \ref{fig:osc}, we get a spatial representation of the oscillation along the loop. The KE$_x$ inside the loop starts quickly building up after $t\sim 38$ minutes, with a spatial distribution resembling the fundamental kink mode of an oscillating loop. After a peak around $t=49$ minutes, a gradual drop in the KE$_x$ and the oscillation amplitude is observed, alongside the loop being further displaced by the background flow. This shows that the efficiency of the energy being supplied to our oscillator drops over time for our simulations. However, due to the simplicity of our model, further studies are required in order further explore the intricacies of this mechanism.

In order to identify the spatial and temporal harmonic structure of our oscillator, we plot the power spectral density along the loop and over a wide spectrum of frequencies, in Fig. \ref{fig:freq}. From this height-frequency ($z-f$) diagram we detect a maximum, corresponding to the eigenfrequency of the fundamental kink standing mode of our system, here given in values of $P^{-1}$. On the same panel we can see an increased value of the spectral density at frequencies around zero, which is caused by the background flow and resulting loop displacement that is present in our system. Alongside the main one, we can see an additional maximum, albeit weaker, at around $1.7-1.8$ $P^{-1}$ frequency. This ratio of $1.8$ between these two maxima agrees with the expected analytical value \citep{andries2005,Safari2007} for our model with a scale height of $H=55$ Mm. This ratio is also close to the detected ratio of $1.4$ between the fundamental and second harmonic of loops undergoing decay-less oscillations \citep{duckenfield2018ApJ}, reinforcing our result that this secondary maximum represents the second harmonic. The seismological application of this finding could be used in order to identify the loop characteristics in decay-less oscillations, stressing the need for further studies of this self-oscillatory process.

\begin{figure*}[t]
    \centering
    \resizebox{\hsize}{!}{\includegraphics[trim={0cm 3cm 5.7cm 0cm},clip,scale=0.25]{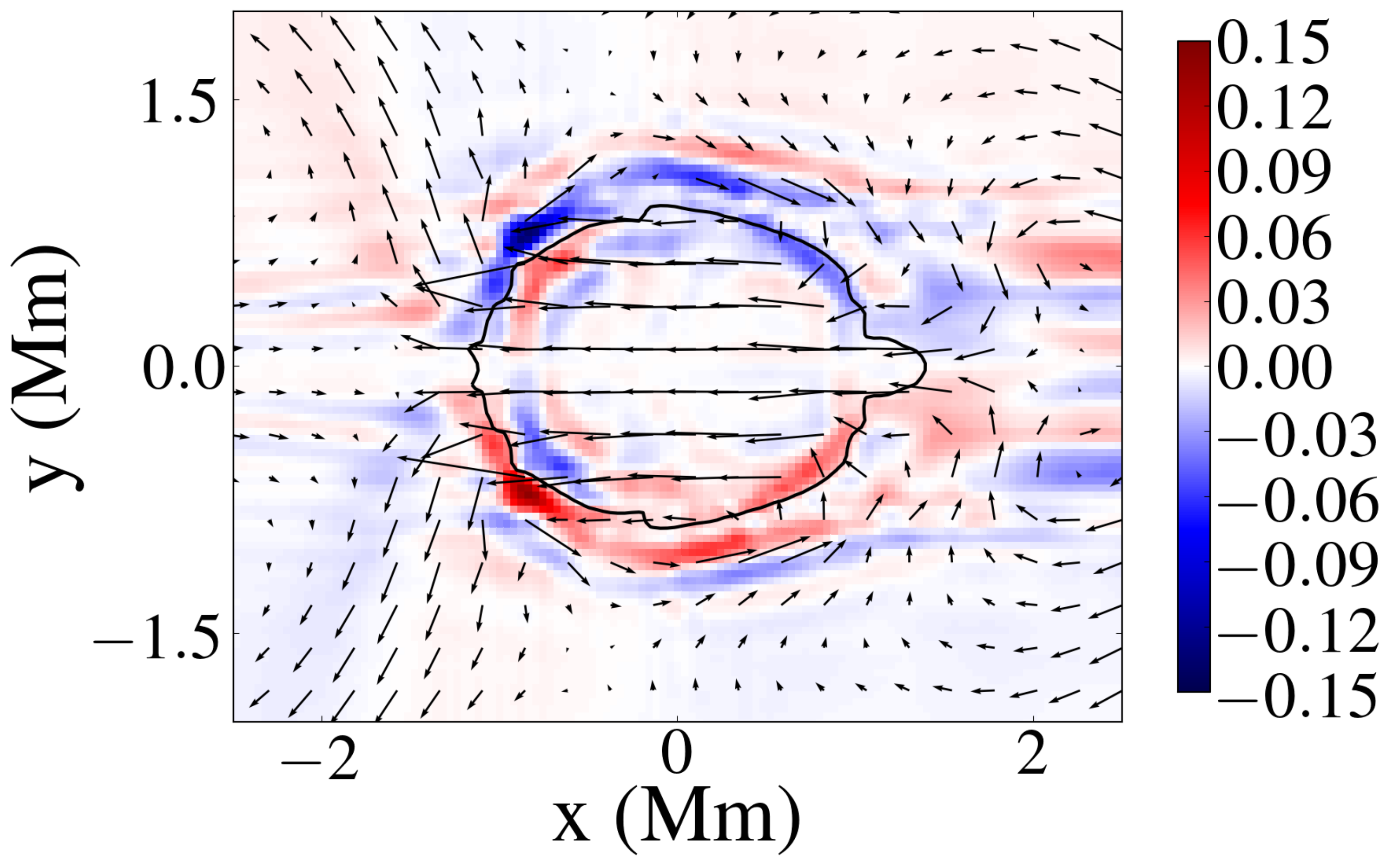}
    \includegraphics[trim={5cm 3cm 5.7cm 0cm},clip,scale=0.25]{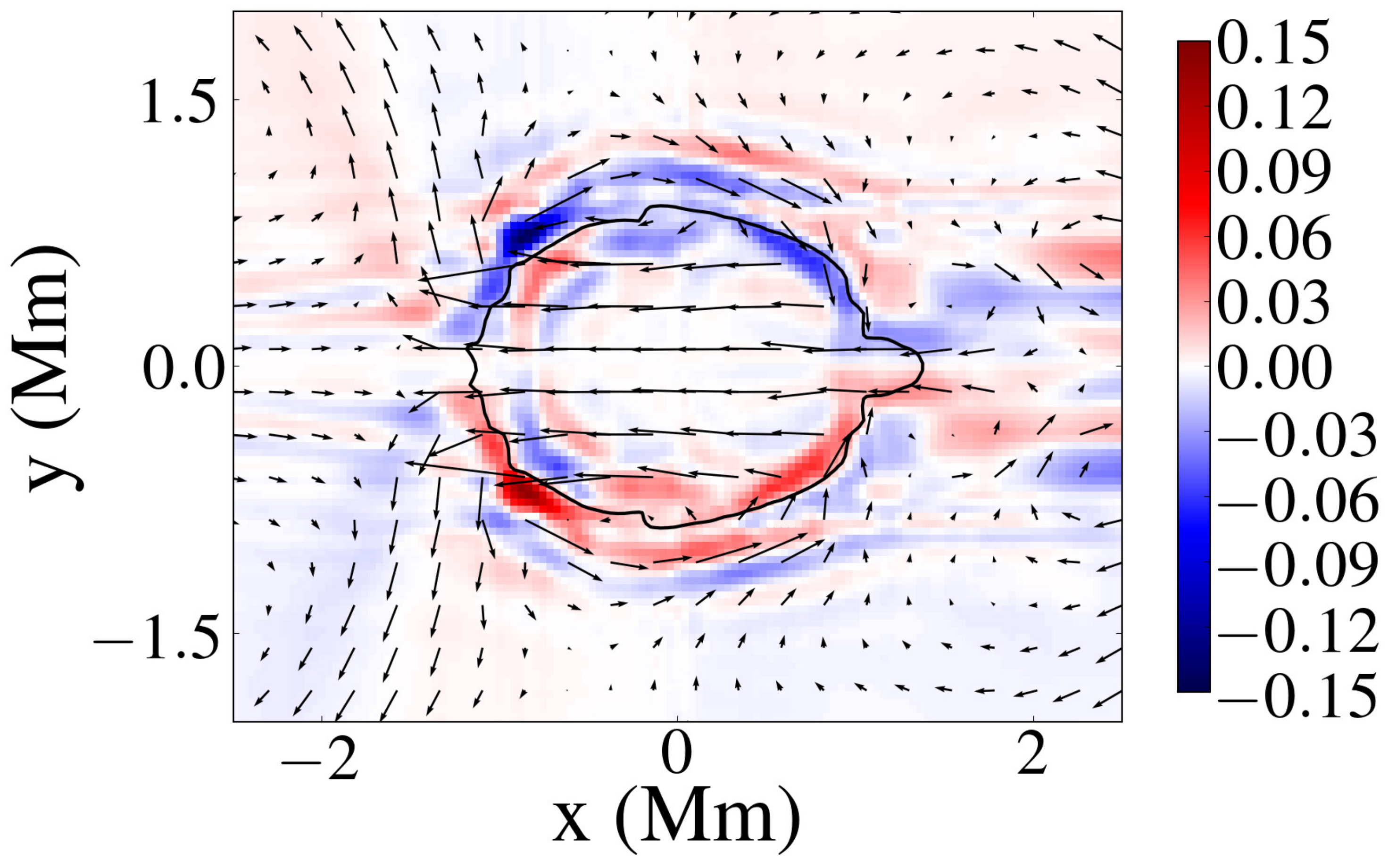}
    \includegraphics[trim={5cm 3cm 5.7cm 0cm},clip,scale=0.25]{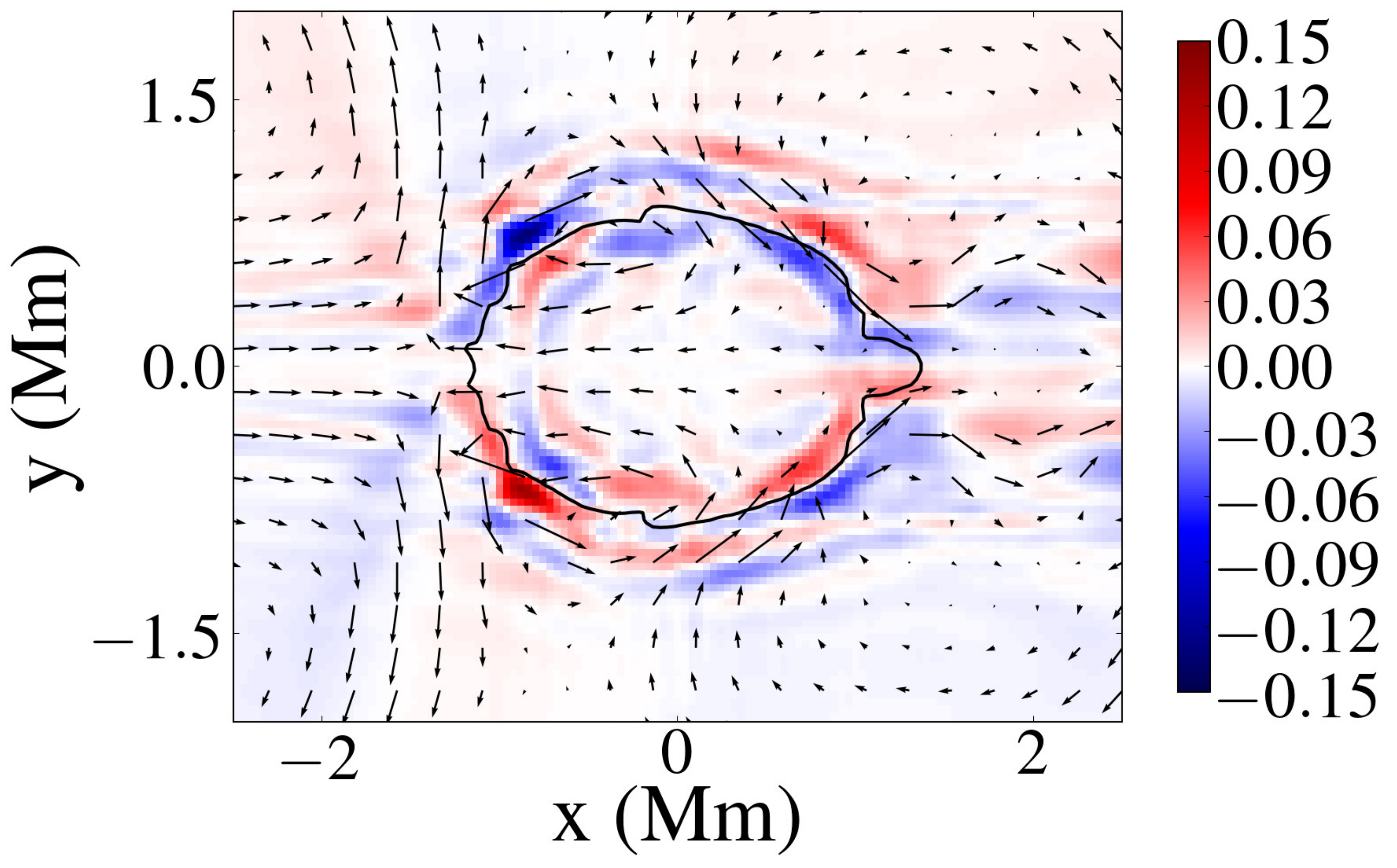}
    \includegraphics[trim={5cm 3cm 0cm 0cm},clip,scale=0.25]{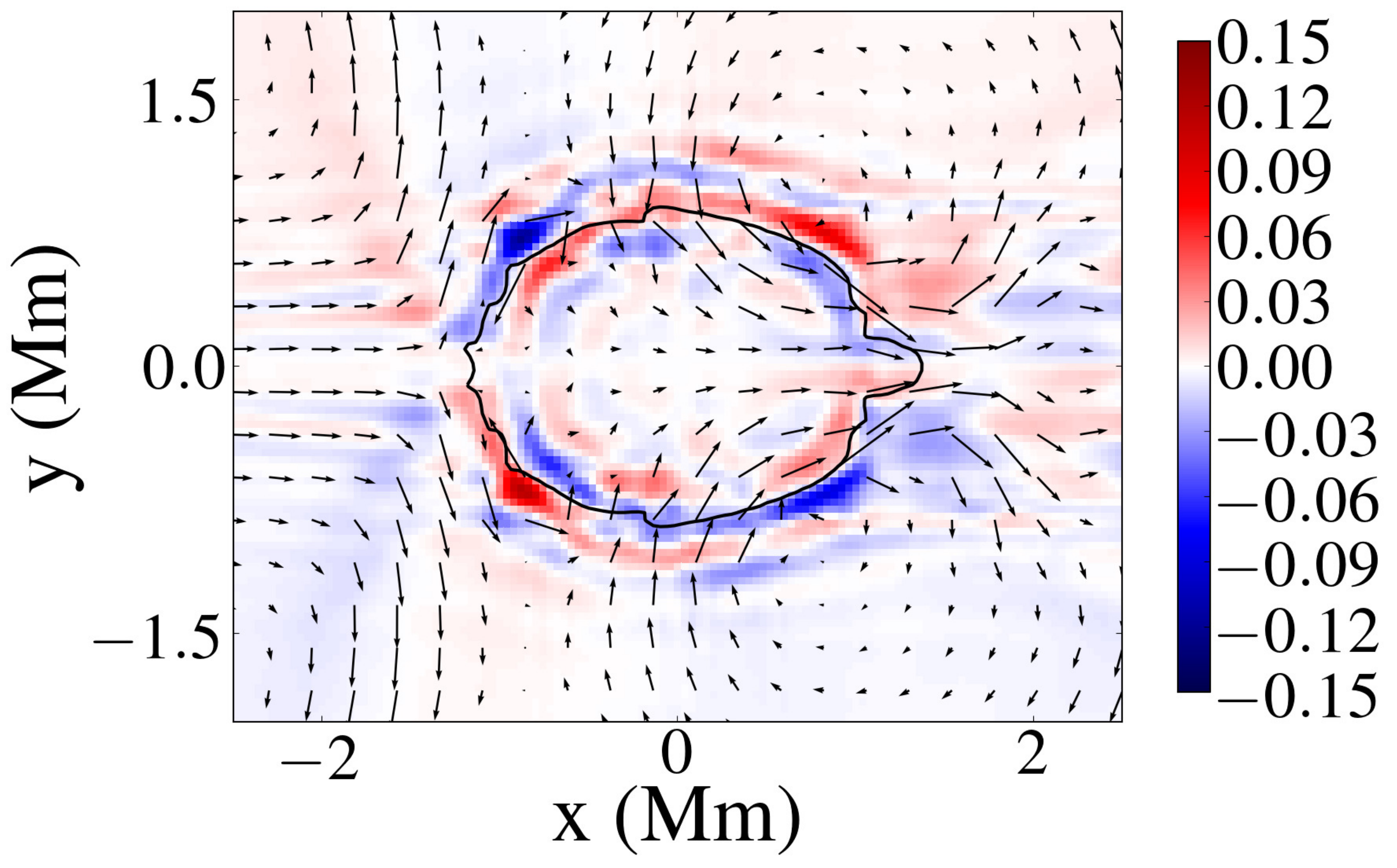}}\\
    \resizebox{\hsize}{!}{\includegraphics[trim={0cm 0cm 5.7cm 0cm},clip,scale=0.25]{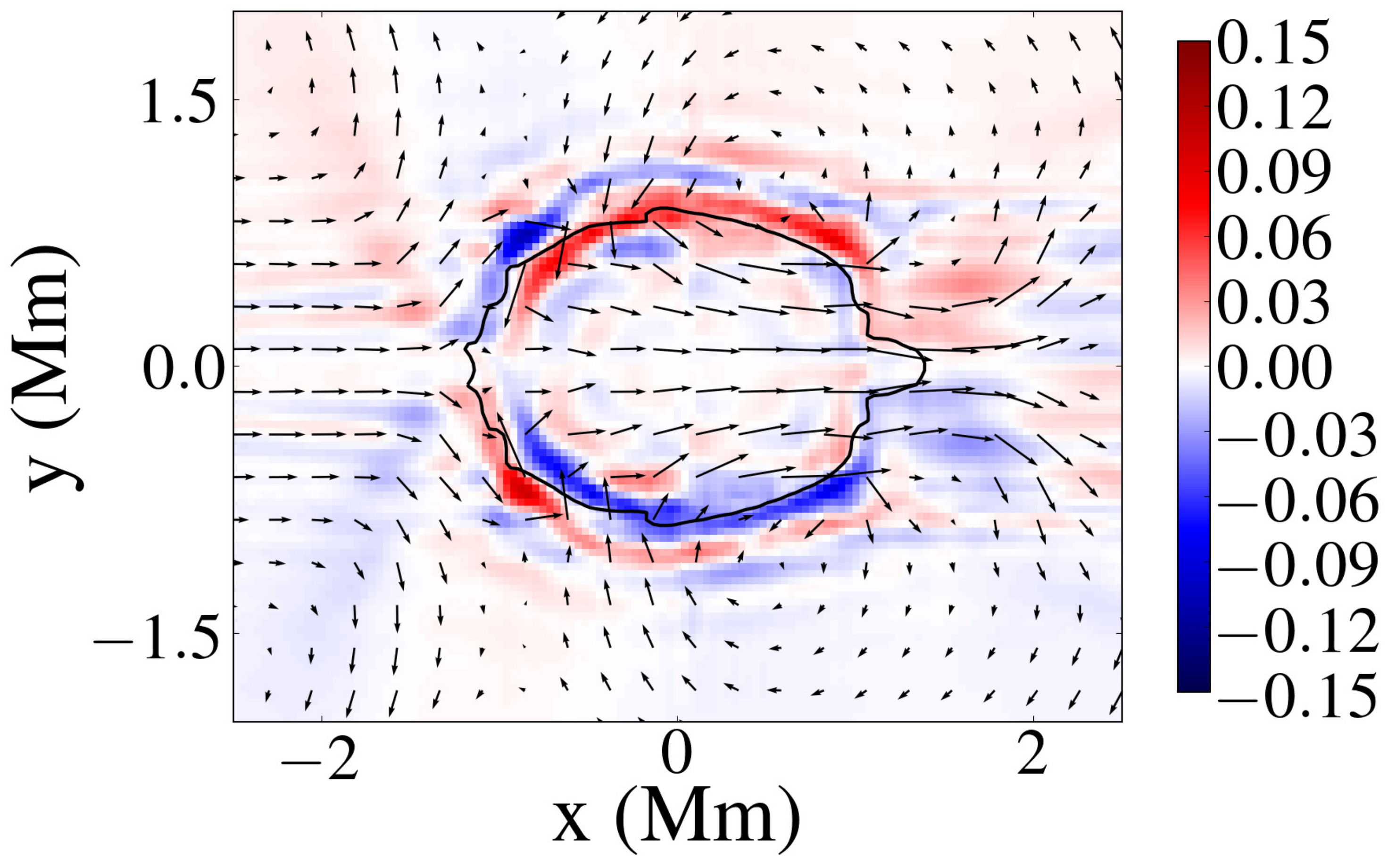}
    \includegraphics[trim={5cm 0cm 5.7cm 0cm},clip,scale=0.25]{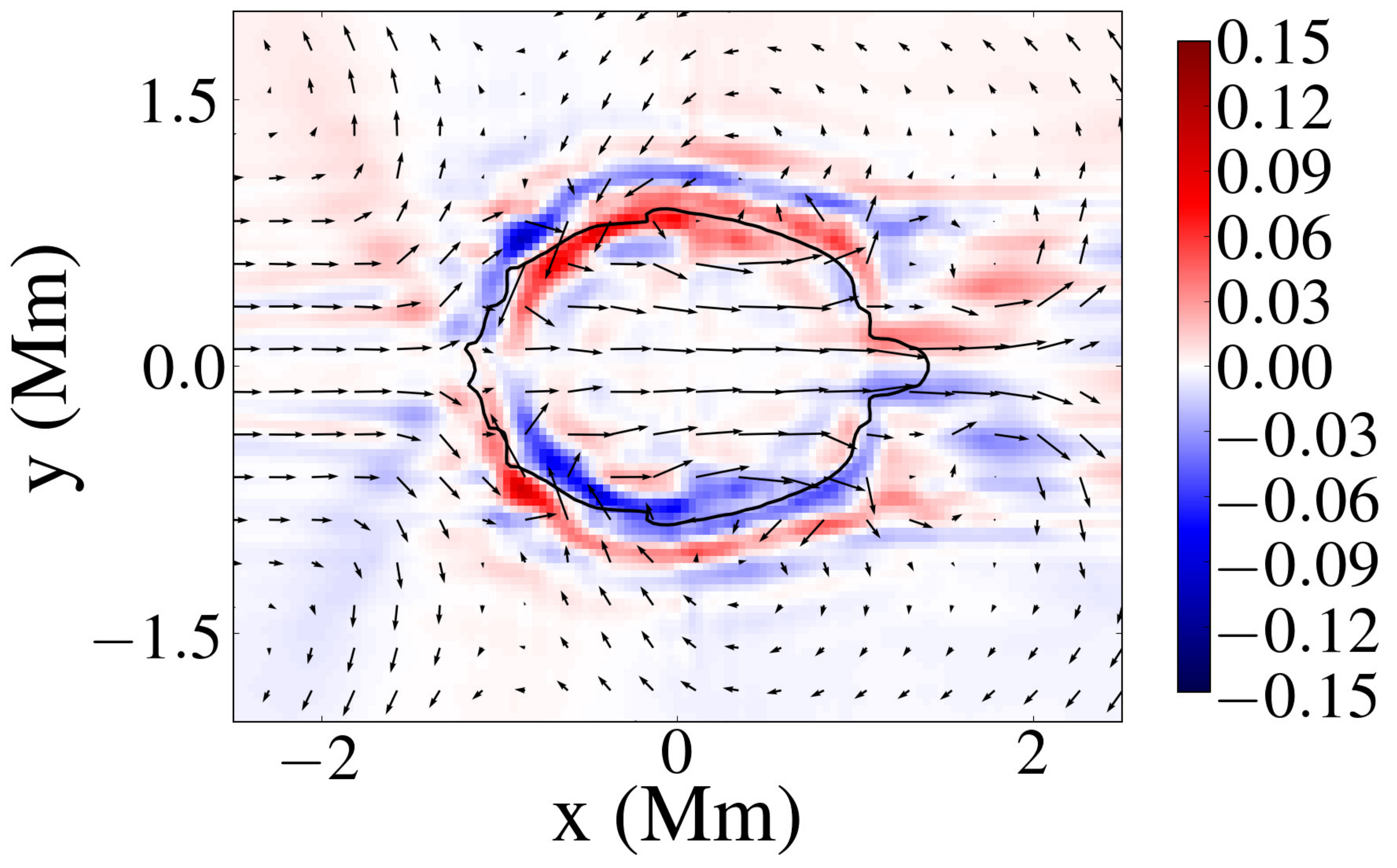}
    \includegraphics[trim={5cm 0cm 5.7cm 0cm},clip,scale=0.25]{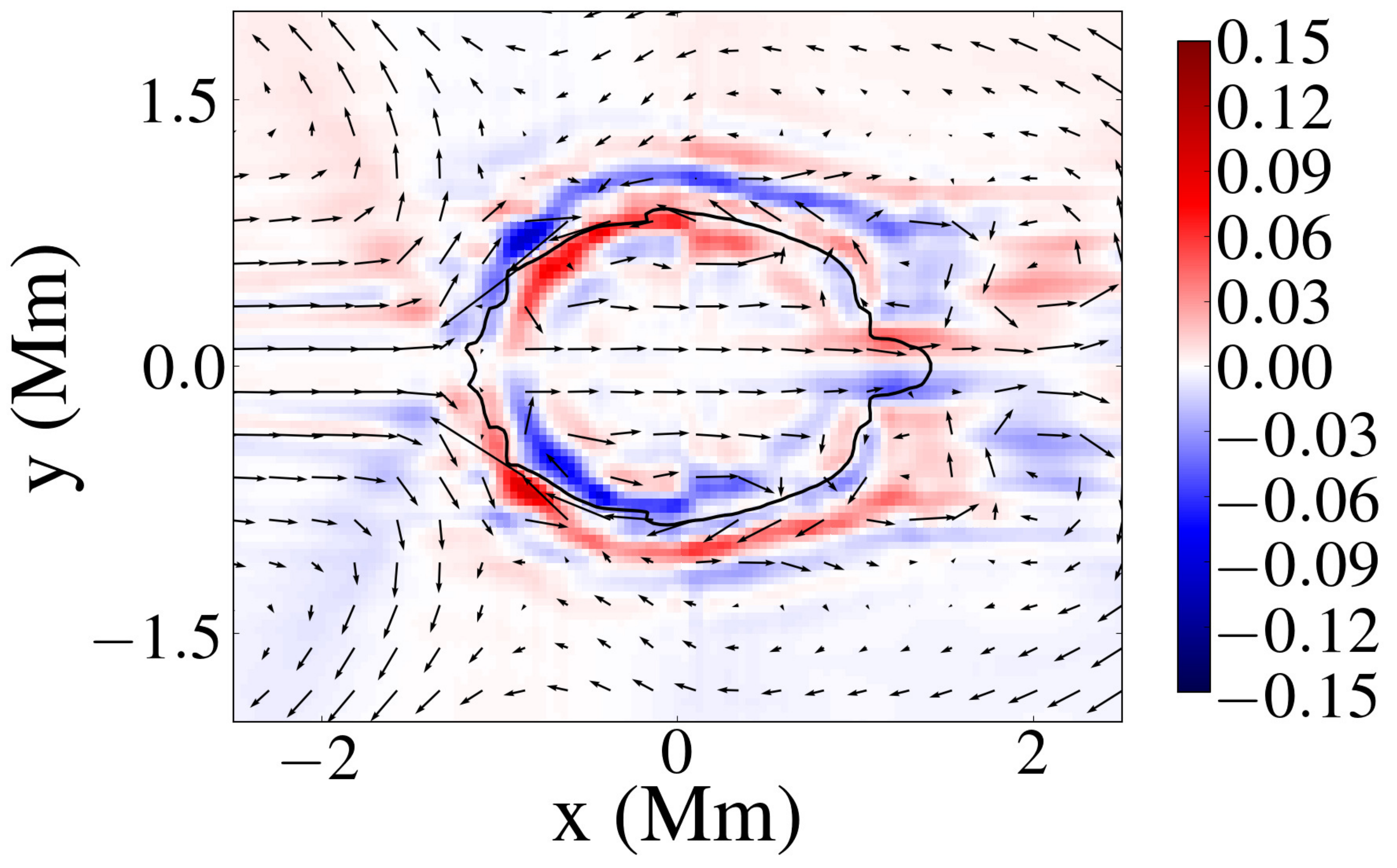}
    \includegraphics[trim={5cm 0cm 0cm 0cm},clip,scale=0.25]{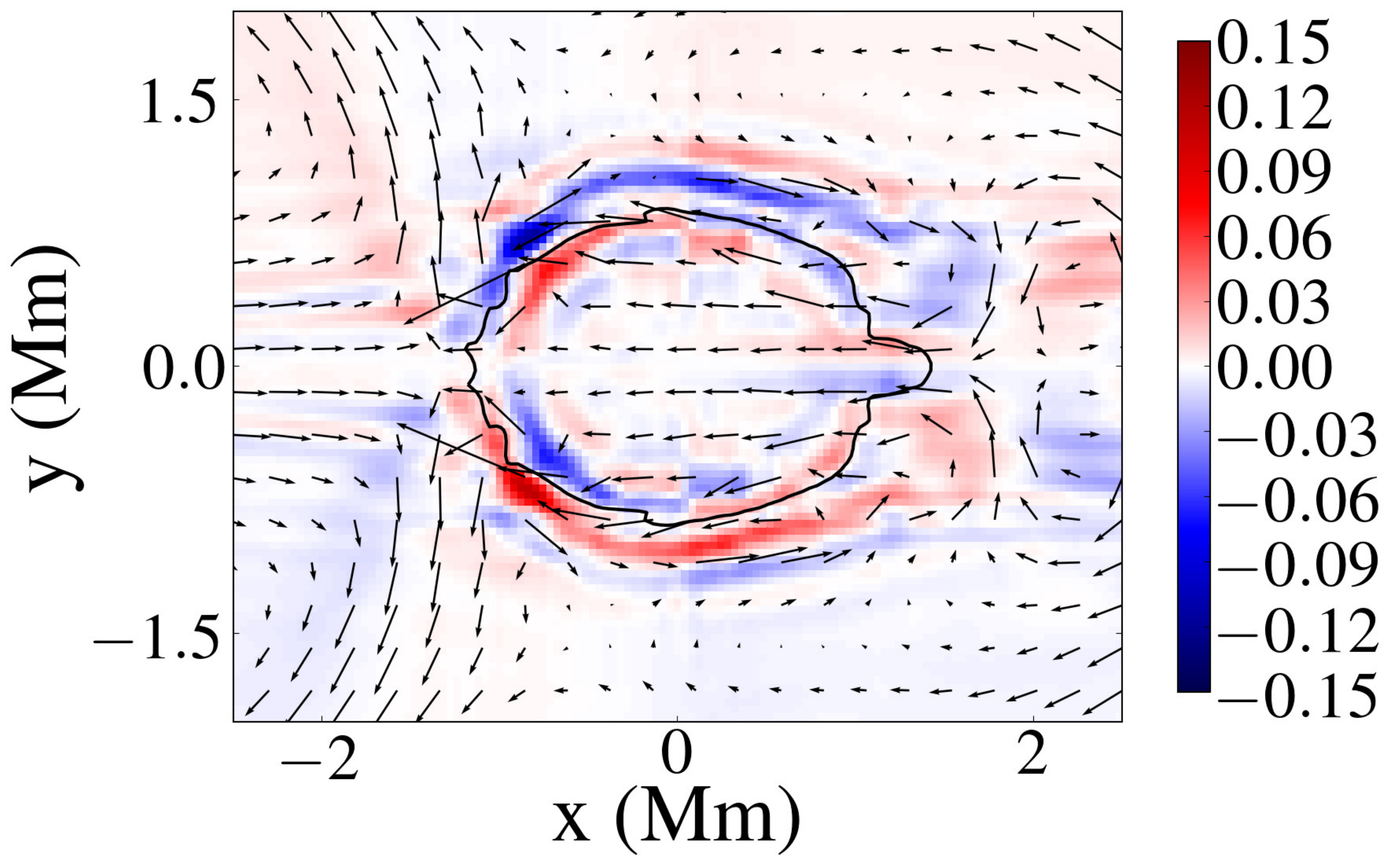}}
    \caption{Contour plots of the plasma $z-$vorticity ($\times\, 0.1285$ Hz) for our oscillating loop at the apex. Overplotted we have a contour of the plasma density ($\rho = 0.6 \times 10^{12}$ kg m$^{-3}$) and the normalized velocity field at the apex for each snapshot. From the upper left to the lower right: snapshots every $20.1$\,s starting at $t = 53.7$ minutes. An animation of these panels is available, covering the oscillation between $t=47.6$ minutes and $t=64.4$ minutes.}
    \label{fig:vfield}
\end{figure*}

As was argued in \citet{nakariakov2016}, an important aspect of having a self-oscillating coronal loop is the presence of a ``slippery" interaction with the background flow, i.e. the ability of the oscillator (here the loop) to be displaced by the background flow before slipping back to its equilibrium position in a periodic manner. This slippery interaction is present in our simulations, and is visualized as vortices forming around and behind the flux tube. We can see this effect taking place in Fig. \ref{fig:vfield} and in its accompanying animation, where cross sections of the loop at the apex are plotted at different times between  $t=19\, P$ and $t=20\, P$ ($P=161$\,s), alongside the velocity field at the same height. From the contours of the $z-$vorticity we can see the drifting of these vortices away from the flux tube, following the background flow. Here we should stress that these vortices are not the driving mechanism behind the oscillations. The oscillation develops through the combined action of a continuous ``push" from the background flow and the magnetic tension of the loop. The vortices are instead connected with the tube ``slipping" through the background flow.

The fact that the vortices are not driving the oscillation can be seen by studying the results of \citet{gruszecki2010} on the phenomenon of Alfv\'{e}nic vortex shedding, studied in 2D for a coronal environment. In that study, the Strouhal number of the flow, defined as
\begin{equation}
    St = \frac{d}{PV_0},
\end{equation}
was found to have values between $0.15$ and $0.25 $. The Strouhal number is a ratio that connects the periodicity of vortex shedding with the characteristics of the obstacle and the flow. For an obstacle of diameter $d=2$\, Mm (like our loop), and a period of vortex generation of $P=161$\,s, a flow of $V_0 \sim 50-80$\,km s$^{-1}$ would be required, for the loop to resonate with and be driven by the vortices. This value is far stronger than the one imposed by the footpoint driver. Instead, the vortices are connected with the tube ``slipping" through the background flow, as we mentioned before. Once the symmetry of the background flow breaks and the process of vortex shedding initiates, the loop then starts to oscillate at its own eigenfrequency, imposing this frequency on the creation of vortices.

\section{Summary and conclusions}

In this study we have explored a mechanism that treats the so-called decay-less oscillations of coronal loops as self-sustained oscillations, which are processes that can turn a nonperiodic driving mechanism into a periodic signal \citep{jenkins2013PhR}. In the simple mechanical model studied in \citet{nakariakov2016}, the energy of the oscillation was provided from a semi-steady constant driving from an external driver with characteristic time scales much longer than the oscillation period. In that paper, supergranulation flows were considered as a possible source of this semi-constant driving of loops, leading to a system resembling a bow on a violin sting \citep{goedbloed1995AIPC}. In our model, the initiation of a weak background flow at one footpoint, eventually led to the start of on oscillation in our gravitationally stratified flux tube. From the power spectral density distribution it is deduced that the oscillation is dominated by the fundamental standing kink mode, while a weaker second harmonic is also revealed. The frequency ratio of the second to the first harmonic is calculated between $1.7$ and $1.8$, similar to the ratio calculated in the observed decay-less oscillations \citep{duckenfield2018ApJ}. The ``slippery" interaction between the loop and the background loop, which is essential for the initiation of the self-oscillation, is detected in our setups as well in the form of vortices around and behind the loop, caused by the oscillation.

However, due to its simplicity, our model only provides a proof-of-concept for a self-sustained oscillation in coronal loops, showing a number of limitations when exploring this mechanism. The formation of the vortices in our setup, which seems to be essential for sustaining the oscillation, has to be further explored in a coronal environment. To that end, expanding the results of \citet{gruszecki2010} about Alfv\'{e}nic vortex shedding to a 3D domain is an essential next step. In addition to that, the oscillation amplitudes that we obtained here by tracking the center of mass, are on the lower side for the observed decay-less oscillations. From the equation of the Rayleigh oscillator used by \citet{nakariakov2016} to describe the self-oscillating process of their mechanical analog, the oscillation amplitude is regulated by a nonlinear term associated with the linear friction between the oscillating mass-spring system and its driver, as well as the driver velocity. Decreasing the eigenfrequency of the system will also lead to an increase in amplitude of the Rayleigh oscillator. Translating that into our setup, we expect the oscillation amplitude to be affected by the energy dissipation mechanisms in our setup, the flow strength and localization, and the characteristics of the loop (e.g. loop length, magnetic field strength and distribution, density-pressure distribution, etc.). A parameter study is therefore required in order to find the connection between the oscillation amplitudes and the physical characteristics of the oscillator and the background flow. This will also be important for better determining the strength of the observed second harmonic, which could be significant for seismology studies using these decay-less oscillations. Moreover, the nature of this background flow must be further explored in more realistic simulations of a coronal loop system, addressing possible external drivers that could cause it, like a flow from supergranulation or an upflow passing the coronal part of a loop \citep{nakariakov2009}. However, in this short study we prove for the first time that such a self-sustaining  mechanism is possible in a coronal environment beyond cartoon and $0$D models, and it should be further explored as a possible interpretation of undamped transverse waves in coronal loops.

\acknowledgments

K.K. is supported by a postdoctoral mandate from KU Leuven Internal Funds (PDM/2019). T.V.D. was supported by the European Research Council (ERC) under the European Union's Horizon 2020 research and innovation program (grant agreement No. 724326) and the C1 grant TRACESpace of Internal Funds KU Leuven. The computational resources and services used in this work were provided by the VSC (Flemish Supercomputer Center), funded by the Research Foundation Flanders (FWO) and the Flemish Government – department EWI.





\bibliography{paper}{}
\bibliographystyle{aasjournal}



\end{document}